\documentclass[useAMS,usenatbib]{mn2e}
\usepackage{color}
\pdfoutput=1

\def\ltsima{$\; \buildrel < \over \sim \;$}
\def\lta{\lower.5ex\hbox{\ltsima}}
\def\gtsima{$\; \buildrel > \over \sim \;$}
\def\simgt{\lower.5ex\hbox{\gtsima}}

\usepackage{natbib}
\usepackage{graphicx} 

\title[The z $\sim0$ star formation distribution function]{The Star Formation Rate Distribution Function of the Local Universe}
\author[M. S. Bothwell et al.]
{M. S. Bothwell$^{1}$\thanks{E-mail:
bothwell@ast.cam.ac.uk},
R. C. Kennicutt$^{1}$, 
B. D. Johnson$^{1}$, 
Y. Wu$^{2}$, 
J. C. Lee$^3$, \newauthor
D. Dale$^{4}$,
C. Engelbracht$^{5}$,
D. Calzetti$^{6}$,
E. Skillman$^7$\\
$^{1}$ Institute of Astronomy, University of Cambridge, Cambridge, CB3 0HA \\
$^{2}$ Infrared Processing and Analysis Center, California Institute of Technology, 1200 E. California Blvd, Pasadena, CA 91125 \\
$^{3}$ Carnegie Observatories of Washington, Pasadena, CA, USA \\
$^{4}$ Department of Physics \& Astronomy, University of Wyoming, Laramie, WY 82071 \\
$^{5}$ Steward Observatory, University of Arizona, Tucson, AZ 85721 \\
$^{6}$ Department of Astronomy, University of Massachusetts, Amherst, MA 01003, USA \\ 
$^{7}$  Department of Astronomy, School of Physics and Astronomy, University of Minnesota, 116 Church Street S.E., Minneapolis,\\
 \hspace{4pt} MN 55455, USA \\
}

\begin{document}

\date{Accepted xxxx. Received xxxx in original form xxxx}

\pagerange{\pageref{firstpage}--\pageref{lastpage}} \pubyear{2002}

\maketitle

\label{firstpage}

\begin{abstract}
We present total infrared (IR) and ultraviolet (UV) luminosity functions derived from large representative samples of galaxies at $z \sim 0$, selected at IR and UV wavelengths from the IRAS IIFSCz catalogue, and the GALEX AIS respectively. We augment these with deep Spitzer and GALEX imaging of galaxies in the 11 Mpc Local Volume Legacy Survey (LVL), allowing us to extend these luminosity functions to lower luminosities ($\sim 10^6 \mathrm{L}_{\sun}$), and providing good constraints on the slope of the luminosity function at the extreme faint end for the first time. Using conventional star formation prescriptions, we generate from our data the SFR distribution function for the local Universe. We find that it has a Schechter form, that the faint-end slope has a constant value (to the limits of our data) of $\alpha=-1.51\pm0.08$, and the `characteristic' SFR $\psi^*$ is $9.2\;\mathrm{M}_{\sun}\; \mathrm{yr}^{-1}$. We also show the distribution function of the SFR volume density; we then use this to calculate a value for the total SFR volume density at $z \sim 0$ of $0.025\pm0.0016\;\mathrm{M}_{\sun}\; \mathrm{yr}^{-1}\; \mathrm{Mpc}^{-3}$, of which $\sim 20\%$ is occurring in starbursts. Decomposing the total star formation by infrared luminosity, it can be seen that $9\pm1 \%$ is due to LIRGs, and $0.7\pm 0.2\%$ is occuring in ULIRGs. By comparing UV and IR emission for galaxies in our sample, we also calculate the fraction of star formation occurring in dust obscured environments, and examine the distribution of dusty star formation: we find a very shallow slope at the highly extincted end, which may be attributable to line of sight orientation effects as well as conventional internal extinction.

\end{abstract}

\section{Introduction}

Measuring the distribution of star formation rates (SFRs) at $z=0$ provides a fundamental observational reference point for galaxy evolution models, being a directly testable result of the hierarchical galaxy formation process. As such, obtaining good constraints on the local star formation rate distribution function is an integral part of calibrating and constraining cosmological models concerning the growth and buildup of structure in the Universe. 

In an ideal world, it would be possible to directly measure the bolometric output of \textit{all} galaxies in some large volume-limited sample, thus measuring - without biases - the true underlying distribution of star formation and luminosity. In reality, however, there are two major constraints to be overcome in order to approach the bias-free distribution functions. Firstly, for large samples the bolometric output and star formation rate must be extrapolated from one or more luminosity components, typically H$\alpha$, UV, or IR, which necessarily involves some uncertainty (particularly for extrapolations based on a single continuum wavelength - see, for example, \citealt{Calzetti:2000aa}). Secondly, the sample's selection function plays a large part in determining the behaviour of the derived distribution, with samples selected in the UV and IR probing very different populations with very different underlying star formation behaviour and dust content (e.g. \citealt{2005ApJ...619L..51B}, \citealt{2006ApJS..164...38I}). The behaviour of dust is also of great interest, as extinction, correlating strongly with star formation, plays a major role in the derived star formation properties of a sample.

Measuring the luminosity function and the SFR distribution function has been an important goal for many years, with much work being dedicated towards successfully estimating the local luminosity function (\citealt{2003ApJ...587L..89T}, \citealt{2005ApJ...619L..47S}, \citealt{2005ApJ...619L..15W}, \citealt{2010MNRAS.401...35W}), the star formation rate distribution (e.g. \citealt{2002NewA....7..337B}, \citealt{2005ApJ...619L..59M}, \citealt{2007ApJS..173..415M}), and the closely related total star formation rate volume density (\citealt{1995ApJ...455L...1G}, \citealt{1998ApJ...495..691T}, \citealt{2004ApJ...615..209H}). These studies, often based on the large homogenous IRAS and GALEX survey datasets, have provided good constraints at the bright, highly star forming end of the distribution. However, the behaviour of the faint end slope is still somewhat unexplored, with most studies suffering significant incompleteness at low luminosities ($\lta10^{8-9} \;\mathrm{L}_{\sun}$) and SFRs ($\lta10^{-1} \;\mathrm{M}_{\sun}\; \mathrm{yr}^{-1}$). These less well-explored populations at the faint end, being numerically dominant, provide a vital insight into the hierarchical galaxy formation process.

It is a well established result that the low-mass slope of the predicted Dark Matter (DM) mass function is steeper than the observed faint-end slope of the luminosity function (LF): this discrepancy is often framed as the well known `missing satellite problem' (\citealt{1999ApJ...522...82K}; \citealt{Moore:1999aa}). The faint-end slope of the luminosity function is a way of quantifying the number of faint galaxies, and while models adopting a Press-Schechter formalism predict a steep faint-end slope of $\alpha \sim -1.8$ (\citealt{1974ApJ...187..425P}; \citealt{2009MNRAS.398.1150B}), observational results suggest a significantly shallower value (\citealt{1999ApJ...521...50M}; \citealt{2003MNRAS.344..307L}).

However, the form of the observed LF is several steps removed from the DM mass function, being the end result of many non-linear baryonic processes. A multitude of physical processes dictate how DM haloes collect and retain their gas, setting the form of the \textsc{Hi} mass function which, through star formation, produces the luminosity function. Measuring and comparing the distribution functions pertaining to these baryonic processes (i.e. the the hydrogen mass function, the star formation rate function, and the luminosity function) has the potential to shed light on the possible reasons that baryons are under-abundant in low mass dark matter haloes, causing the shallow slope and `missing satellites'.

To attempt to obtain good constraints on the extreme faint end, we use data from the Local Volume Legacy survey, a volume-limited dataset of galaxies within 11 Mpc with deep multi-band Spitzer, optical, and GALEX imaging. Being volume limited, this survey provides an excellent and unbiased database from which to calculate IR and UV luminosities and star formation rates, which can be combined with the same parameters from the larger datasets. For the first time, therefore, we are able to augment our large survey datasets with deep imaging of this statistically complete sample of local galaxies, and extend the local luminosity and star formation rate distributions functions 1-2 orders of magnitude deeper. 

In this paper, we present statistically complete SFR distribution functions and luminosity functions which extend 1-2 orders of magnitude deeper than previous studies. The data provide the first ever SFR distribution study complete at both the bright and extreme faint ends, and give an unbiased view of the distribution of star formation and structure at $z\sim 0$. We use our samples to accurately constrain the true value - and breakdown - of the star formation rate volume density in the $z=0$ Universe. In order to investigate the role of extinction (which correlates strongly with SFR), we also use IR and UV photometry to analyse the behaviour of dust obscured star formation (measured via the ratio $\mathrm{L}_{\mathrm{IR}} / \mathrm{L}_{\mathrm{FUV}}$), and present `extinction distribution functions', analogous to the luminosity function.

Throughout this paper we assume a concordance $\Lambda$-CDM cosmology with parameters (H$_0, \; \Omega_m, \; \Omega_{\Lambda}) =  (72 \;\mathrm{km s}^{-1}\; \mathrm{Mpc}^{-1}$, 0.27, 0.73). The Local Volume sample, for which peculiar motions overwhelm the Hubble flow, has distances based on a number of direct distance estimators wherever possible (see \S{\it 2.2.2}).

\section[]{Methodology}
\subsection{Sample Selection}
\label{sec:samples}
As mentioned in \S1 above, for any observationally-defined sample the nature of the selection function will go a long way to determining the physical properties of the sample. Any non-volume-limited sample used to calculate a luminosity function must be strictly flux limited at some wavelength, so that when weightings are applied (for example volume weightings, as per the $1/\mathrm{V}_{\mathrm{max}}$ method) the resulting function, $\Phi(\mathrm{L})$, becomes statistically representative of the underlying population as a whole (an inconsistent flux limit would lead to bias through inconsistent volume corrections). 

The wavelength at which this limit is imposed will to a large extent determine the properties of the sample. An IR-selected sample (selected at $60\mu$m for example) which is then crossed with a UV database will contain more dusty, obscured systems than an initially FUV-selected sample which is subsequently cross matched with an IR database. \cite{2007ApJS..173..404B} demonstrated that the bolometric luminosity functions for IR and UV selected samples diverged for luminous systems ($\mathrm{L}_{\mathrm{bol}} > 5 \times 10^{10}\; \mathrm{L}_{\sun}$), with a significant proportion of the massive, luminous systems detected strongly at 60$\mu$m being non-detected in the FUV due to their extreme dust attenuation. At the low-luminosity end, samples suffer from the opposite bias; late-type, metal poor, star forming galaxies may lack sufficient dust to re-process the UV photons into IR emission, and as a result can be weak- or non-detections at 60$\mu$m despite having measurable levels of ongoing star formation (\citealt{2009ApJ...703..517D};  \citealt{Calzetti:2010aa}).

Clearly, defining a sample in terms of selection at a single wavelength and extrapolating global properties will lead to an inaccurate view of the behaviour of star formation and extinction in the overall population. In this paper, we therefore use separate datasets selected at 60$\mu$m and GALEX FUV - hereafter, these will be referred to as the `IR-selected' and `UV-selected' samples respectively. These dual samples can then be combined, and the resultant dataset encapsulates the full range of star forming behaviour, ensuring that both obscured and unobscured star formation are represented. 

\subsubsection{IR-selected sample}
For our IR-selected sample, we make use of the Imperial IRAS Faint Source Catalogue redshift database (IIFSCz) compiled from the IRAS Faint Source Catalogue (FSC) by  \cite{2009MNRAS.398..109W}. This is a large (61\% of the sky) database consisting of 60,303 galaxies selected at 60$\mu$m. For our analysis, we require that all galaxies have redshifts, and we include both spectroscopic and photometric redshifts in our compilation (\citealt{2009MNRAS.398..109W} discuss in detail the accuracy of their photometric redshifts, derived using neural network fitting, concluding that they are accurate to 1.5-2.3\%) A total of 44,622 galaxies in the IIFSCz have a secure redshift identification. We then cross matched this subsample against the GALEX AIS, resulting in a total of 25,768 galaxies observed with both IR and UV: this comprises our IR-selected sample. Naturally, some of the galaxies observed with GALEX were not detected due to a low UV flux. We distinguished between GALEX non-detections (i.e. upper limits) and galaxies never observed with GALEX by crossmatching with a $0.6^{\circ}$ search radius, equivalent to the size of a GALEX tile. 

To ensure that redshift evolution effects played no significant part in the luminosity distribution, we applied a $z<0.1$ cut to our sample (which will slightly reduce the number density of the most luminous systems, which are under-represented in the local Universe). We also applied a 60$\mu$m flux limit of 0.36 Jy, which was then used to define the value of V$_{\mathrm{max}}$ for each galaxy in the IR sample - this limit corresponds to the 90\% completeness limit of the IRAS FSC, and is higher than the formal 60$\mu$m flux limit of the FSC of 0.2 Jy . We also applied a minimum redshift cut of $z = 0.005$ ($\sim$20 Mpc), to remove any galaxies for which peculiar motion would overwhelm the Hubble flow, leading to highly uncertain distances based upon recession velocity alone. 

This last cut in effect removes IR-faint (L$_{\mathrm{IR}} \leq 10^9 \;\mathrm{L}_{\sun}$) galaxies from the sample; the volume-limited LVL sample samples this region of luminosity space well however, so no information is lost overall. This near field cut also has the effect of eliminating much of the GALEX `shredding' problem whereby nearby extended sources are resolved as multiple objects, resulting in artificially lowered UV fluxes (see, e.g, \citealt{2007tS..173..267S}). By comparing the UV fluxes of our IR-selected sample with those of Buat et al. (2007), who used their own photometric extraction technique, we estimate that at most  $\sim5\%$ of our galaxies suffer from photometric extraction issues (such as shredding). Constructing luminosity functions consisting solely of the subset of galaxies common to both samples (302 members), we find that the UV LF produced by our fluxes is essentially identical to that built from independently-obtained UV fluxes. As such, we proceed with the assumption that our UV fluxes are robust. 

The final IR sample, after applying all the above cuts, consists of 10,252 galaxies. 

To confirm that working with this subsample of the IIFSCz does not bias our results, we constructed the (monochromatic) 60$\mu$m luminosity function for our subsample, and compared them to the equivalent LFs of \cite{2010MNRAS.401...35W} which were constructed using the complete sample; the LF derived from our subsample does not differ significantly from those derived from the parent sample (see Fig. 1 for the LF of our original, `unadulterated' IR sample). The main difference is a sharper cut-off at the most luminous end of the luminosity function for our sample, which is attributable to the high-$z$ cut - the parent sample, having no such cut, includes many ULIRGs and HyLIRGs (systems with L$_{\mathrm{IR}} > 10^{12} \; \mathrm{L}_{\sun}$ and $10^{13} \;\mathrm{L}_{\sun}$ respectively), which are very rare in the $z\sim0$ Universe.

\subsubsection{UV-selected sample}
The UV-selected sample was taken from the paper by \cite{2007ApJS..173..404B}, who assembled a UV-selected sample of galaxies to assess the relative contributions of obscured and unobscured star formation in the local Universe. To briefly summarise, their sample was selected by applying a UV cut of FUV = 17.5 mag to the GALEX All-Sky Imaging Survey (AIS) catalogue, and cross-matching with the IRAS PSCz in areas uncontaminated by foreground (i.e. galactic) cirrus emission. The resulting effective area is 2210 deg$^2$ - while smaller than the large area covered by the IIFSCz (which, at 61\% of the sky, covers over an order of magnitude more sky area), this is still substantial and will avoid the clustering biases and sensitivity to small-scale inhomogeneities that are the weakness of `pencil beam'-type surveys. Distances for the UV sample galaxies were obtained from NED.

The same distance cuts as the IR-selected sample were applied ($0.005 < z < 0.1$). The low- and high-$z$ cut-offs remove 8 and 4 galaxies respectively from the parent sample of 606 -- our UV-selected sample therefore consists of 595 galaxies.

\subsubsection{The Local Volume sample}


To augment our IR- and UV-selected samples in the low luminosity regime, we use data on a complete sample 
of nearby galaxies collected by the GALEX 11HUGS
(11 Mpc H$\alpha$ UV Galaxy Survey) and Spitzer LVL 
(Local Volume Legacy) programs.  The sample is 
dominated by dwarf galaxies, and is thus ideal for 
studying the nature of systems with low SFRs, low 
metallicities and low dust contents.  UV, and mid- to 
far-IR flux catalogs are published in Lee et al. (2010) 
and Dale et al. (2010), respectively.  Details on the
sample selection, observations, photometry are provided in 
those papers, and in Kennicutt et al. (2008), who describe 
the overall parent 11 Mpc sample and H$\alpha$ imaging survey.
A brief summary of the dataset is given here.

The total parent Local Volume sample contains 436 objects.
Galaxies are compiled from existing catalogs
(as described in Kennicutt et al. 2008), and the selection
is divided into two components. The primary component
of the sample aims to be as complete as possible
in its inclusion of known nearby star-forming galaxies
within given limits.  It consists of spirals and irregulars
brighter than B = 15 mag within 11 Mpc that avoid the
Galactic plane ($|b|>$20$^{o}$).  These bounds
represent the ranges within which the original
surveys that provided the bulk of our knowledge
on the Local Volume galaxy population have been shown
to be relatively complete, while still spanning a large 
enough volume to probe a representative cross section 
of star formation properties.
The secondary component of the sample consists of galaxies
that are within 11 Mpc and have available H$\alpha$
flux measurements, but fall outside one of the
limits on brightness, Galactic latitude, or
morphological type.  It is a composite of targets that 
were either observed by Kennicutt et al. (2008) as telescope 
time allowed, or had H$\alpha$ fluxes published in the literature.
Subsequent statistical tests, as functions of
B-band apparent magnitudes and HI fluxes (compiled from
the literature), show that the subset of galaxies 
with $|b|>20^{\circ}$ is relatively complete to 
$\mathrm{M}_{\mathrm{B}} \lta -15$ and $\mathrm{M}_{\mathrm{{HI}}}> 2 \times 10^8$
M$_{\odot}$ at the edge of the 11 Mpc volume 
Lee et al. (2009).

Subsequent GALEX UV imaging primarily targeted
the $|b|>30^{\circ}$, $B<15.5$ subset of the sample.
The more restrictive latitude limit was imposed to
avoid excessive Galactic extinction and fields
with bright foreground stars and/or high background
levels for which observations would be prohibited due
to GALEX's brightness safety restrictions.  Deep,
single orbit ($\sim$1500 sec) imaging was obtained
for each galaxy, following the strategy of the
GALEX Nearby Galaxy Survey \citep{Gil-de-Paz:2007aa}.  GALEX
observations for a significant fraction of the remaining
galaxies beyond these limits were also taken by other GI
programs.
Overall, GALEX data are available for $\sim90$\% of the 
436 galaxy sample.

Finally, Spitzer IRAC mid-infrared and MIPS far-infrared
imaging was also obtained for the $|b|>30^{\circ}$, $B<15.5$
subset of the sample through the Local Volume Legacy program.
This sub-sample with both UV and IR coverage contains 257 
galaxies and is used in the following analysis.

The resulting dataset provides an unprecedented multi-wavelength view of star formation in the nearby Universe. In particular, the multi-band Spitzer IR observations allow for bolometric luminosities to be calculated (using the algorithms provided by \citealt{2002ApJ...576..159D}) in a manner consistent with the IRAS fluxes available for our larger datasets.

\subsection{Constructing the Luminosity Functions}

\subsubsection{1/V$_{\mathrm{max}}$ method}

The `classic' way of constructing a luminosity function is based on the estimator

\[
\Phi(L) dL = \sum_i \frac{dL}{\mathrm{V}_{\mathrm{max}}(L_i, S^{lim}_{\nu})} 
\]

Where V$_{\mathrm{max}}$ is the volume enclosed by the maximum distance at which galaxy $i$ would be observable, given the flux limits of the survey ($S^{lim}_{\nu}$ - defined by the IR and UV respectively for the IR- and UV-selected samples, and the B-band for the LVL sample) and the galaxy's luminosity - this `maximum volume' is then used to weight each galaxies' contribution to the final function $\Phi$(L). This inverse volume weighting is designed to counteract the Malmquist-type bias which would be encountered by a pure number counting exercise: the volume probed by a flux limited survey varies as a function of luminosity, and as such faint galaxies (which are only seen nearby) are under-represented; conversely, bright galaxies seen out to large distances are over-represented. Weighting by 1/V$_{\mathrm{max}}$ eliminates this effect, and reconstructs the true underlying luminosity distribution. This particular form of the weighting (as opposed to first binning into luminosity bins, then applying a mean V$_{\mathrm{max}}$ to all galaxies within the bin) reduces error resulting from binning of the data - in essence, each galaxy is assigned its own bin of width dL. 

The 1/V$_{\mathrm{max}}$ method has the great advantage of being a non-parametric estimator, as it does not assume any form of $\Phi$(L). However, it does suffer from two weaknesses: firstly, binning in luminosity space is sometimes required, which necessarily involves some loss of information, and the resulting form of $\Phi$(L) can be somewhat sensitive to the particular binning used. Secondly, and more seriously, it is highly sensitive to local density enhancements (see \citealt{1988MNRAS.232..431E} for a detailed discussion). The Local Volume represents a significant overdensity compared to the cosmic mean, which will manifest in a flux limited survey as an enhancement at the faint end of the luminosity function. \cite{2004AJ....127.2031K} compared the B-band luminosity density in the local 8 Mpc and find that it is 1.7-2.0 times the global luminosity density (as derived from both the Sloan Digital Sky Survey and the Millennium Galaxy Catalogue). We therefore correct the local density downwards by a factor $1.85 \pm 0.15$, and incorporate this uncertainty in the error on the derived faint end slope. This is consistent with the values found by \cite{2009ApJ...706..599L} and \cite{2010arXiv1009.4705L} by comparing the LVL and the the field \textsc{Hi} mass function (the Local Volume is over-dense by a factor of 1.4), the B-band LF (a factor of 2.3) and the UV LF (a factor of 2).

\subsubsection{Maximum likelihood method}

The weakness to density fluctuations can be overcome by adopting a parametric estimator, which assumes the form of the luminosity function is universal (with the precise shape being determined by some free parameters), which allows the density to be factored out. One such statistic is the `maximum likelihood' method, as described by (e.g.) \cite{1988MNRAS.232..431E} and \cite{1991ApJ...372..380Y}. 

We take a \cite{1976ApJ...203..297S} function to be the assumed form of the luminosity function: 
\[
\Phi(L) dL = \Phi^* \left(\frac{L}{L^*}\right)^{\alpha} e^{-(L/L^*)} d\left(\frac{L}{L^*}\right).
\]
When fitting this to data using a maximum likelihood method, the free parameters are obtained by maximising the value of $\Delta$ with respect to $\alpha$ and $L^*$,
\[
\Delta = \sum_i \log F_i,
\]
where
\[
F_i = \frac{\Phi_i}{\Psi} =   \frac{(L_i/L^*)^{\alpha} e^{-(L_i/L^*)}} {\Gamma(\alpha+1, L_{lim}/L^*)},
\]
and where $\Psi = \int \Phi(L) dL$ is the cumulative luminosity function (equal to $\Gamma$, the normal incomplete gamma function for a Schechter LF), and $L_{lim}$ is the minimum luminosity at which the sample is complete. Essentially, a grid of Schechter functions with varying parameters is explored, and each is assigned a `likelihood' using the data. Our adopted function is taken to be the function which maximises the `likelihood'.  

While this method has the aforementioned advantage that it is insensitive to local density fluctuations, it has the disadvantage of being similarly insensitive to the absolute normalisation of the luminosity function (note that $\Phi^*$ cancels out in the expression for $F_i$). This can be recovered using number counting, as the integral under the derived best fit function is just the number of galaxies:
\[
N(>L_{lim}) = \int_{L_{lim}}^\infty \Phi(L') dL \;=\Phi^* \; \Gamma(\alpha+1, L_{lim}/L^*).
\]

The uncertainty on the parameters of the luminosity function - derived using either method - can be difficult to estimate. A major source of potential error, particularly at the faint end, is uncertainty on the distance. \cite{2008ApJS..178..247K} discuss the distance estimates for the Local Volume galaxies in detail: direct distance estimators are used wherever possible (Cepheids, red giant branch, Tully-Fisher relation, supernovae), and when a distance from one of these methods is not available distances are estimated from the Hubble velocity, corrected for a local group dipole effect as derived by \cite{1996AJ....111..794K}. The uncertainties on the distance (both random, and systematic to the flow model) are typically $7 - 15 \%$.

The effect of this distance uncertainty (and the effect of the photometric flux uncertainty) will be to shift galaxies into neighbouring bins of luminosity (or SFR). As a result, the errors on the various luminosity function parameters are highly correlated, and have a complex dependence on the errors in the raw data. We deal with these errors using a Monte Carlo resampling method, by creating a large number of realisations of the luminosity function drawn - with replacement - from the parent sample. Parameters with well defined errors (i.e. flux) were allowed to vary randomly according to a Gaussian likelihood function defined by the 1$\sigma$ parameter error, and the standard deviation in the resultant luminosity functions in each bin was taken to be an estimate of the error.



\section{Analysis}

\subsection{Removing the AGN contribution}
\label{agn}

\begin{figure}
 \centerline{\includegraphics[scale=0.55, clip=true, trim = 10mm 125mm 40mm 30mm]{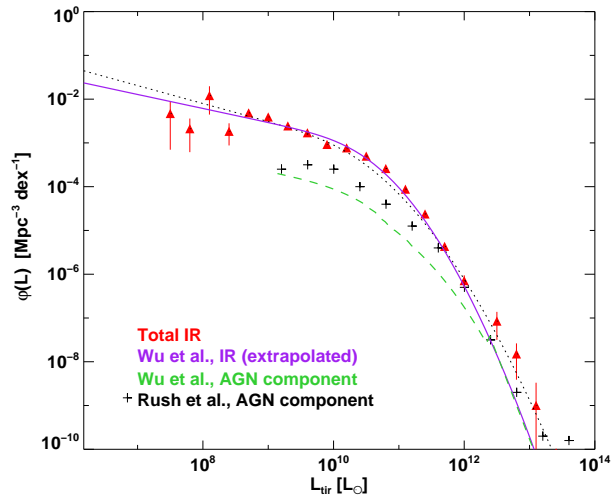}}
 \caption{IR LF for our original IR-selected sample (red triangles), plotted with an extrapolated IR LF from Wu et al. (purple line). The AGN contribution to Wu et al.'s LF is shown as a green line. Plotted as black crosses is a previous estimate of the AGN contribution to the IR LF, from Rush et al. (1993). For reference, the canonical IRAS IR LF (Takeuchi et al. 2003) is plotted as a dotted black line. Note that our original IR-selected sample is shown here before any cuts, and is therefore identical to the LF of Wang and Rowen-Robinson (2010).}
 \label{fig:agn_lf}
\end{figure}


Obscured Active Galactic Nuclei (AGN), are a potential source of contamination for our sample, if we want to interpret the emission in terms of pure star formation. Unified models of AGN assume a dusty torus - which emits strongly in the IR - surrounding a central engine; this bright IR emission can be falsely interpreted as originating from dust heated by star forming regions, which leads to an overestimation of the star formation rate of the galaxy. If we are to accurately estimate the star formation rate density of the volume containing our sample, it is important to remove such sources of possible contamination.


AGN number amongst the most extreme objects in the Universe, and, when active, have prodigious bolometric outputs. The effect of AGN subtraction will, therefore, be to reduce the very upper end of the luminosity function, while leaving the behaviour at $\mathrm{L}<10^{\sim10 - 11}\; \mathrm{L}_{\sun}$ relatively unchanged (see e.g. \citealt{2010ApJ...709..884Y}; \citealt{2010MNRAS.402.1693H}). At the upper end, star formation occurs almost entirely in highly extincted, dusty environments (\citealt{Calzetti:2010aa} and references therein), where IR is by far the best SFR tracer: we therefore consider the effect of AGN contamination on our IR-selected sample only, as the UV-selected and LVL datasets will have a negligible AGN contribution \citep{2007tS..173..267S}.

It is difficult to perfectly remove all AGN on a galaxy-by-galaxy basis without detailed spectroscopy being available for the entire sample, using which the AGN component can be separated from the star forming component (see e.g. \citealt{2010arXiv1008.2932F}). 
In the absence of spectral data for our entire sample, a better method is to remove the AGN component statistically from the sample as a whole. This requires a knowledge of the global behaviour of the AGN population, in the form of an AGN luminosity function which can then be subtracted from our original `total' luminosity function, leaving only the star forming component.


To this end, we use the data in the work by \cite{2010b_Wu}, a spectroscopic study of a subset of the 24um selected 5mJy Unbiased Spitzer Extragalactic Survey (5MUSES) sample \cite{2010arXiv1009.1633W}. This subset focused on galaxies with $z<0.3$ from 5MUSES, with $<z>\sim0.12$. The 226 objects in this subsample were analysed with mid- and far-IR photometry (allowing a full characterisation of their IR SED), and mid-IR spectroscopy - allowing each source's luminosity to be decomposed into AGN and star-forming components. For the purpose of this analysis, we will examine the IR luminosity function of their sample and use the relations given in their work to derive the contribution to the IR LF from AGN. We then take this fractional AGN contribution (as a function of IR luminosity) and subtract it from our data, leaving us with IR data for our galaxies which can be interpreted in terms of pure star formation.

Figure \ref{fig:agn_lf}  shows the luminosity function for our IR-selected sample. Also plotted is the total IR luminosity function for the galaxies reported in \cite{2010arXiv1009.1633W}, derived by taking their 15$\mu m$ LF and transforming into an IR LF using the average luminosity conversion given. The AGN component of their IR LF is also shown. As expected, their SED fitting shows that the bolometric IR luminosity for AGN sources has a larger contribution from shorter wavelengths than for star forming sources: they calculate $\log (\mathrm{L}_{14\mu m}/\mathrm{L}_{\mathrm{IR}}) = -1.19 \pm 0.10$ for star forming sources, but just $-0.57 \pm 0.19$ for AGN dominated systems.

Being IR selected, Wu et al.'s sample has the advantage of having similar properties to our own, and therefore we expect that the AGN contribution, calculated as a function of IR luminosity, will be applicable to our data. The good agreement between the two IR LFs -- our own and Wu et al.'s -- confirms this, and suggests that the AGN fraction derived from their sample is indeed applicable to our own. We attribute the slight bright-end discrepancy between the functions to small number uncertainty at the upper end of their sample - Wu et al.'s complete sample consists of 330 galaxies, and will sample the rarer IR-bright galaxies poorly. As a qualitative check, we compare the AGN contribution derived here to a previous result, the AGN luminosity function at 12$\mu m$ as calculated by \cite{1993ApJS...89....1R}.

As discussed above, the AGN contribution to the star formation distribution only becomes significant at the bright end ($>\sim 10^{12} \mathrm{L}_{\sun}$), with the contribution from AGN being $<10\%$ for non-LIRGs, with star formation rates below $\sim 15 \; \mathrm{M}_{\sun}\; \mathrm{yr}^{-1}$. Both measures of AGN activity agree well, with a small (factor of $\sim 2$) discrepancy at the low ($<10^{11} \mathrm{L}_{\sun}$) end - the low value of $\Phi$(AGN) at these luminosities renders this difference relatively unimportant. We do note, however, that the 5MUSES sample has spectral AGN estimators which are more accurate than the photometric estimators of the older work, and it is to this that we attribute the difference. 

All luminosity functions, star formation distributions, and related values calculated hereafter have had the above AGN LF contribution subtracted as per the estimate from Wu et al., and are therefore interpretable in terms of pure star formation.

\subsection{Luminosity functions}

\begin{figure}
 \centerline{\includegraphics[scale=0.55, clip=true, trim = 10mm 125mm 40mm 30mm]{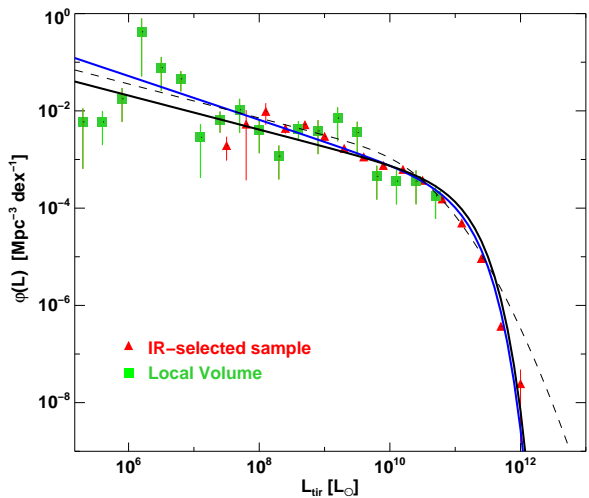}}
 \caption{The bolometric IR luminosity function for the IR sample. 1/V$_{\mathrm{max}}$-derived points are shown for the IR-selected sample (red triangles) and the Local Volume galaxies (green squares). The black line is a least squares Schechter fit to these points, the blue line is the maximum likelihood Schechter function fit. Errors (1$\sigma$) were calculated using Monte Carlo bootstrapping. The dashed black line is the `canonical' IRAS LF, from Takeuchi et al. (2003).}
 \label{fig:tir_lf}
\end{figure}

\begin{figure}
 \centerline{\includegraphics[scale=0.55, clip=true, trim = 10mm 125mm 40mm 30mm]{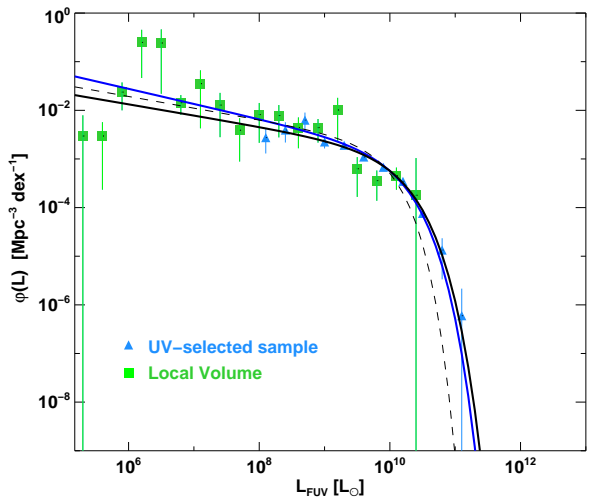}}
 \caption{The FUV luminosity function for the UV sample. 1/V$_{\mathrm{max}}$-derived points are shown for the UV-selected sample (blue triangles) and the Local Volume galaxies (green squares). The black line is a least squares Schechter fit to these points, the blue line is the maximum likelihood Schechter function fit. Errors (1$\sigma$) were calculated using Monte Carlo bootstrapping. The black dashed line is the UV LF from Wyder et al. (2005).}
 \label{fig:uv_lf}
\end{figure}

All galaxies in our datasets have been observed with multiple bands in the IR: either with Spitzer (at 24$\mu$m/70$\mu$m/160$\mu$m), or IRAS (12$\mu$m/25$\mu$m/60$\mu$m/100$\mu$m). As such, very accurate estimates - to better than 1\% for most galaxies - of their bolometric IR flux (defined as the integrated flux from 8$\mu$m-1000$\mu$m) can be made, using the three-component prescriptions provided by  \cite{2002ApJ...576..159D}. 

Fig. \ref{fig:tir_lf} shows the TIR luminosity function for both the large IR sample (red triangles) and the LVL data (green squares). As we are interested in measuring the true underlying distributions, rather than just the local density enhancement, the number densities for the LVL data in Fig. \ref{fig:tir_lf} (and for all LVL volume densities hereafter) have been adjusted downwards by a factor of 1.85 as discussed above. The data points were derived using the 1/V$_{\mathrm{max}}$ method, and the black line is the best fitting Schechter function, obtained using standard least squares fitting. The blue fit to the data is the Schechter function obtained using maximum likelihood fitting, as discussed above. The two methods agree on all counts, with the exception of a small discrepancy in the faint end slope ($1.41 \pm 0.09$ for the 1/V$_{\mathrm{max}}$ method, $1.53 \pm 0.08$ for maximum likelihood). Power law fits to the IR data, once the AGN
component was removed, were generally poor - in all cases, a Schechter function was able to fit the data more closely. 

For comparison, the luminosity function for IRAS IR-selected galaxies from \cite{2003ApJ...587L..89T} (corrected slightly for our cosmological parameters) has been overplotted as a dashed line. There is a discrepancy at the high end, caused by both AGN removal (c.f. our original sample shown in Fig \ref{fig:agn_lf}) and the redshift cut (taking $z<0.1$ reduces the number of rare bright objects). Indeed, Takeuchi et al. attribute the `bright end bump' seen in both our samples to the increasingly dominant AGN contribution at luminosities $\mathrm{L}_{\mathrm{IR}} > 5.0 \times 10^{11} \mathrm{L}_{\sun}$ (the point at which their power law LF diverges from our Schechter LF), which we have removed as per \S \ref{agn}. Apart from this difference, the LF of the parent sample is essentially identical to our UV-matched subsample, suggesting that we have not introduced undue bias by insisting on a UV observation for inclusion in the sample (see \S \ref{non} below).  

Fig. \ref{fig:uv_lf} shows the FUV luminosity function for the large UV sample (blue triangles) and the LVL data (green squares). The fitted black and blue curves are again, respectively, the least-squares fit to the 1/V$_{\mathrm{max}}$ data points, and the maximum likelihood fit. Being the sample from Buat et al. (2007), the UV LF is identical to theirs. The dashed line is the field FUV LF from GALEX, presented by \cite{2005ApJ...619L..15W} (again corrected for the more recent cosmology). Again, the close similarity between our sample and the canonical `field' LF suggested that our data do provide a representative sample of star forming galaxies in the local Universe. We note however that there is a discrepancy with the LF of Wyder et al. (2005) at the UV-bright end. Again, the two LF derivation methods produce similar fits - the faint end slope is $1.23 \pm 0.09$ for the 1/V$_{\mathrm{max}}$ method, $1.31 \pm 0.09$ for maximum likelihood). 

\subsubsection{Non-detections, upper limits, and incompleteness considerations}
\label{non}
The construction of all the luminosity functions and star formation distribution functions was carried out both by removing all galaxies with only upper flux limits (UV flux limits in the IR-selected sample, and vice-versa), and by treating the upper flux limit as a detection equal to the limiting flux magnitude - as in \cite{2005ApJ...619L..59M}, the results in all cases do not depend on whether the galaxies are included, so we opted to include them. 

This insensitivity to the inclusion of non-detections does not extend to the distribution of obscured star formation, as traced by the ratio $\mathrm{L}_{\mathrm{IR}} / \mathrm{L}_{\mathrm{FUV}}$. This parameter is highly sensitive to low flux values, such as those assumed from an upper limit (being a ratio, low UV fluxes lead to high values of $\mathrm{L}_{\mathrm{IR}} / \mathrm{L}_{\mathrm{FUV}}$ ), and is discussed more in \S\ref{sec:irx}.

We also have to consider the effect of sample incompleteness on the form of our star formation rate function. The large IR and UV datasets will suffer incompleteness starting at $\sim 0.1\, \mathrm{M}_{\sun}\; \mathrm{yr}^{-1}$ (note the downturns the the IR and UV samples shown in Figs. \ref{fig:tir_lf} and \ref{fig:uv_lf} starting at $\sim 10^8 \mathrm{L}_{\sun}$). This is, however, the star formation rate at which the LVL data are highly complete, and hence our resultant function will remain statistically robust here. 

At the faintest end of our function, however, we run into LVL incompleteness which cannot be ignored. \cite{2009ApJ...692.1305L} calculate the LVL dataset to be complete to $\log \, \mathrm{SFR} = -2.5\, \mathrm{M}_{\sun}\; \mathrm{yr}^{-1}$ based on the resultant H$\alpha$ flux limits of the sample. Below this level, too, small number statistics 
lead to derived star formation rates being somewhat uncertain. We do not truncate our sample below this SFR, but it must be noted that values of $\Phi$ calculated at the faint end are inherently uncertain, and for reference we mark this regime on our star formation function shown in Fig. \ref{fig:sfr_tot}. We note, however, that calculations of the faint end slope do not depend on this uncertain regime, and all derived parameters do not change if we fit only to the statistically complete data.

It is worth noting that the volumes probed by our field IR and UV datasets are large enough to avoid bias due to large scale structure. We can define a `correlation volume', derived from the correlation length, which is an estimate of the absolute minimum volume you need to probe in order to be reasonably free of bias due to structure. We estimate this to be 500 Mpc$^3$ (approximating the correlation radius as 5 Mpc; \citealt{Zehavi:2005aa}). For a flux limited survey, the volume probed will vary as a function of luminosity; at a luminosity of $10^8 \;\mathrm{L}_{\sun}$ -- the lower limit of our field data -- our IR and UV datasets respectively probe volumes of 20 and 80 times this minimum `correlation volume'. 

\subsection{The distribution of star formation}
 \label{sec:sfr}

For each galaxy in our sample, we derived a star formation rate as follows. The IRX, defined as
\[
\mathrm{IRX} = \log\left( \frac{\mathrm{L(TIR)}}{\mathrm{L(FUV)}_{\mathrm{obs}}}\right)
\]
is used as a measure of the internal dust absorption. This then can be used to accurately estimate the UV attenuation; it is known that the value of IRX is a good tracer of UV attenuation, and -- importantly -- remains robust independent of the details behind the extinction, such as geometry and dust properties (see \citealt{Meurer:1999aa}). Conversions between IRX and A(FUV) in the literature are given by \cite{Calzetti:2000aa}; \cite{2004MNRAS.349..769K}; \cite{2005ApJ...619L..51B}; and \cite{Cortese:2008aa}. For the purposes of this paper, we use the IRX--A(FUV) relation derived by \cite{2005MNRAS.360.1413B} :
\[
A(FUV) = -0.028 x^3 + 0.392 x^2 + 1.094 x + 0.546,
\]
where $x$ is the IRX defined above.

Once the observed UV luminosity has been corrected for dust attenuation using the above expression, we converted to a star formation rate as per the method described by \cite{2006ApJS..164...38I}:
\[
\log \mathrm{SFR} \; [\mathrm{M}_{\sun} \; \mathrm{yr}^{-1}] = \log (\mathrm{L}_{\mathrm{FUV, corr}}\; [\mathrm{L}_{\sun}]) - 9.51
\]

This calibration uses a Salpeter IMF from 0.1 to 100 M$_{\sun}$, was derived specifically for the GALEX bands, and is similar to other estimators in the literature. For comparison the \cite{1998ARA&A..36..189K} conversion factor: 
\[
\mathrm{SFR} \; [\mathrm{M}_{\sun} \; \mathrm{yr}^{-1}] = 1.4 \times 10^{-28} \; \mathrm{L}_{\nu} \;[\mathrm{ergs}\, \mathrm{s}^{-1}\,\mathrm{Hz}^{-1}]
\]
when converted from a monochromatic UV luminosity - taking the effective central wavelength of the GALEX FUV filter to be 1532 \AA\ - gives $ \log \mathrm{SFR} \; [\mathrm{M}_{\sun} \; \mathrm{yr}^{-1}] = \log (\mathrm{L}_{\mathrm{FUV, corr}}\; [\mathrm{L}_{\sun}]) - 9.56$: $\sim 12\%$ lower than our value. We adopt the former conversion, as it was derived specifically for the GALEX filters by \cite{2006ApJS..164...38I}. We note that this method does carry some implicit assumptions, such as a constant star formation history, and a lack of stochasticity in populating the upper (UV-emitting) regions of the IMF. As noted above, for the purpose of calculating SFRs we include non detections, by assigning a flux equal to the detection limit -- the overall results do not change if we instead discard all non-detected galaxies. 


\begin{figure}
 \centerline{\includegraphics[scale=0.55, clip=true, trim = 10mm 125mm 40mm 30mm]{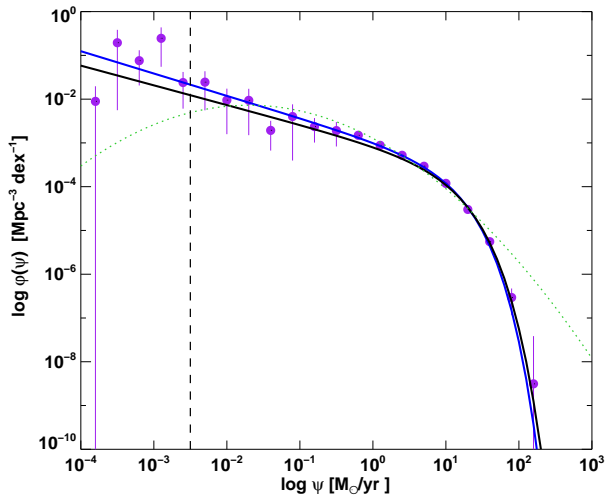}}
 \caption{The star formation rate distribution function for the resultant combined sample, as described in \S\ref{sec:sfr}. The black line is a least squares Schechter function fit to these points, the blue line is the maximum likelihood Schechter function fit. The vertical dashed line is drawn at $\log\, \mathrm{SFR} = -2.5 \;\mathrm{M}_{\sun} \; \mathrm{yr}^{-1}$, the level at which incompleteness becomes significant.  The green log-normal function indicates the SFR function given by Martin et al. (2005). Errors (1$\sigma$) were calculated using Monte Carlo bootstrapping.}
 \label{fig:sfr_tot}
\end{figure}

We then generated a `resultant' sample from our large UV and IR datasets using the `incoherent combination of domain independent samples' method (as described by  \citealt{1980ApJ...235..694A}). Using this method, both samples can be considered together: the combined sample is split into two `domains', defined by $1: [S_{IR} > S^{lim}_{IR}$ and $S_{UV} > S^{lim}_{UV}$], and 2: [$S_{IR} < S^{lim}_{IR}$ and $S_{UV} > S^{lim}_{UV}$].  Any duplicates must be dealt with, as each galaxy is clearly just a single probe of the population. The duplicate is removed, and the remaining galaxy is assigned whichever value of V$_{\mathrm{max}}$ (i.e. based on either the IR or UV limit) is \textit{larger} - this is because for the object to be included, it is sufficient that it satisfies the weaker selection criteria. Objects in the first domain then have V$_{\mathrm{max}}$ values calculated as normal. Objects in the second domain, being constrained by having an IR flux below the limit of the IR-selected sample (and thus only appearing in the UV selected sample), have a reduced region of parameter space to exist in, and have their values of V$_{\mathrm{max}}$ defined by (V$_{\mathrm{max, IR}}$ - V$_{\mathrm{max, UV}}$). 

We cannot apply the same technique to combine with the Local Volume data, as the LVL sample is designed to be volume limited and therefore has a less cleanly-defined selection function; where the samples overlap with LVL, we simply weight the two datasets inversely by their bootstrap-derived errors. The final, combined sample consists of 10,704 galaxies (257 from the LVL, 562 from the UV, and 10,141 from the IR, where duplicate galaxies have been removed as explained above).



The star formation rate function is shown in  \ref{fig:sfr_tot}. The purple data points are V$_{\mathrm{max}}$-derived points for the combined sample as described above. We use least squares fitting to fit a standard Schechter function to the points: the best fitting parameters are found to be ($\Phi^* = 0.00015\pm0.0003\; \mathrm{Mpc}^{-3},\; \psi^*=9.0\pm0.3\;\mathrm{M}_{\sun} \; \mathrm{yr}^{-1}, \;\alpha=-1.48\pm0.07$). This Schechter function is shown in Fig. \ref{fig:sfr_tot} as a black line.

We also used the maximum likelihood method described above to find the best-fitting Schechter function to the combined sample; the best fitting parameters (are found to be ($\Phi^* = 0.00016\pm0.0004\; \mathrm{Mpc}^{-3},\; \psi^*=9.2\pm0.3\;\mathrm{M}_{\sun} \; \mathrm{yr}^{-1}, \;\alpha=-1.51\pm0.08$), and this is plotted on Fig. \ref{fig:sfr_tot} as a blue line. The two methods of calculating the star formation rate function match closely, suggesting that the V$_{\mathrm{max}}$ points are not overly biased by the presence of clustering. Hereafter, we adopt the parameters of the maximum likelihood-derived function. 

The faint end slope of -1.51 continues monotonically almost to the limits of the data, until the low galaxy numbers available at SFRs $<10^{-3}\; \mathrm{M}_{\sun} \; \mathrm{yr}^{-1}$ lead to the degradation of the relation due to noise. This is in contrast to previous extrapolations (e.g. \citealt{2005ApJ...619L..59M}) which predicted a lognormal form for the star formation rate distribution function, with a maximum value of $\Phi(\psi)$ at $\psi \sim 10^{-2} \;\mathrm{M}_{\sun} \; \mathrm{yr}^{-1}$ and a gradual decline thereafter. 

This result is particularly interesting in the light of theoretical predictions of structure formation. There has been a long-running conflict between the small-scale predictions for Dark Matter haloes, and the observational results of the galaxies that dwell inside them \citep{1999ApJ...522...82K}. Galaxy formation models predict a scale-invariant form for Dark Matter, which clusters hierarchically even on the smallest scales; this manifests as a steep faint end slope to the mass function, of $\alpha \sim -1.8$. Observations of galaxy luminosity functions, however, have found a dearth of small scale objects at the bottom of the luminosity function, indicating a much shallower faint end slope. Either the predictions for the Dark Matter mass function are incorrect, or the complex non-linear processes involved in galaxy formation conspire to suppress the formation of baryonic structures on the smallest scales. 

One suggestion has been that the lack of satellite-scale galaxies results from star formation being systematically suppressed on the smallest scales (i.e. \citealt{2000ApJ...539..517B}; \citealt{2005ApJ...632..872R}); the form of our SFR distribution function, however, is consistent with the idea that this is not the case: a monotonically increasing faint-end slope to the SFR distribution function suggests that any process suppressing star formation must be operating in a scale-free manner with respect to halo mass. 


\subsection{The star formation rate volume density}
\begin{figure}
 \centerline{\includegraphics[scale=0.55, clip=true, trim = 10mm 125mm 40mm 30mm]{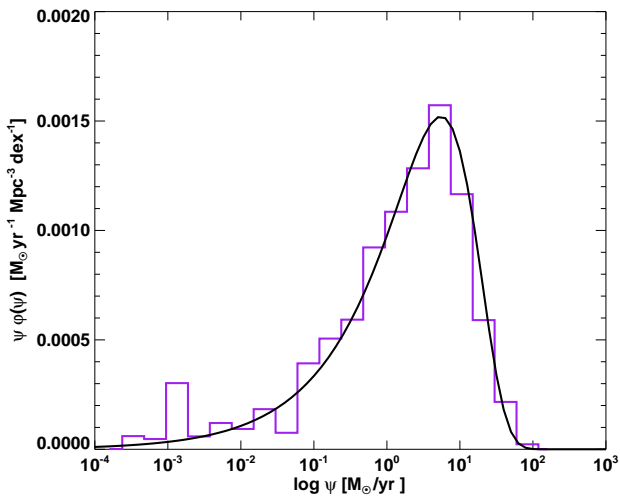}}
 \caption{The star formation rate volume density distribution function, for the resultant combined sample as in Fig. 4. The purple histogram shows the V$_{\mathrm{max}}$-derived data as above, and the black fit to the data is the convolved Schechter function $\psi\, \Phi (\psi)$, with the maximum-likelihood fit parameters as above.}
 \label{fig:sfr_den}
\end{figure}

Using the distribution of star formation rate, it is possible to calculate the distribution of star formation rate \textit{volume density}. We define the distribution function of star formation rate density:
\[
\psi  \,\Phi(\psi) = \psi \; \Phi^* \left(\frac{\psi}{\psi^*}\right)^{\alpha} e^{-(\psi/\psi^*)} d\left(\frac{\psi}{\psi^*}\right).
\]

Fig. \ref{fig:sfr_den} shows the distribution of star formation rate volume density, for the `combined' sample shown in Fig. \ref{fig:sfr_tot}, along with the convolved Schechter fit for the parameters given above. 

We can therefore integrate $\psi  \,\Phi(\psi) $ to give the total star formation rate volume density in the local Universe,

\[
\rho_\mathrm{SFR} = \int_0^\infty  d\psi \; \psi  \,\Phi(\psi),  
\]
which for the parameters of the Schechter fit to our data ($\Phi^* = 0.00016\pm0.0004\; \mathrm{Mpc}^{-3},\; \psi^*=9.2\pm0.3\;\mathrm{M}_{\sun} \; \mathrm{yr}^{-1}, \;\alpha=-1.51\pm0.08$) gives $\rho_\mathrm{SFR} = (25\pm1.7) \times 10^{-3} \;\mathrm{M}_{\sun}\; \mathrm{yr}^{-1}\; \mathrm{Mpc}^{-3}$. 

\begin{table*}
\centering
\begin{minipage}{100mm}
\caption{Derivations from the literature of the star formation rate volume density in the local Universe.}
\begin{tabular}{lllc}
\hline
Reference & SFR tracer  & $<z>$ & SFRD \\
&&&($10^{-3}\; \mathrm{M}_{\sun} \;\mathrm{yr}^{-1}$ Mpc$^{-1}$) \\
 \hline
 \hline
\cite{Gallego:2002aa}& [OII] & $0.025$ & $9.3\pm3$ \\
\cite{2000MNRAS.312..442S}& [OII] & 0.15 & $23\pm3$ \\
\cite{Hogg:1998aa} & [OII] & 0.20 & $11 \pm 4$ \\
\cite{1995ApJ...455L...1G}& H$\alpha$ & $0.022$ & $12\pm5$ \\
\cite{1998ApJ...495..691T}& H$\alpha$ &0.2 & $25\pm 4$\\
\cite{2000MNRAS.312..442S}& H$\alpha$ & 0.15 & $14\pm3$ \\
\cite{2003ApJ...587L..27P} & H$\alpha$ &0.025 & $25\pm 4$\\

\cite{Ly:2007aa}& H$\alpha$ &0.08 & $13\pm 4$\\
\cite{Hanish:2006aa} & H$\alpha$ &0.01 & $16^{+2}_{-4}$\\
\cite{2004MNRAS.351.1151B} & H$\alpha$ & 0.15& $29 \pm 5$ \\
\cite{Dale:2010aa}& H$\alpha$ & 0.16& $10^{+6}_{-4} $ \\
\cite{Westra:2010aa}& H$\alpha$ & 0.05& $6 \pm 2 $ \\
\cite{Westra:2010aa}& H$\alpha$ & 0.15& $12 \pm 3 $ \\

 \cite{2002MNRAS.330..621S}& 1.4 GHz & $0.005$ & $21\pm5$ \\

\cite{1989ApJ...338...13C}& 1.4 GHz & $0.005$ & $21\pm0.5$ \\
\cite{2000MNRAS.312..442S}& FUV & 0.150 & $39\pm5$ \\

\cite{2005ApJ...619L..59M}& FUV+IR & 0.02 & $21\pm 2$\\

This work & FUV+IR & 0.05  &$25 \pm 1.6$ \\

\hline
\end{tabular}
\label{tab1}
\end{minipage}
\end{table*}

This is in good agreement with most recent derivations of this result: see Table \ref{tab1} for a compilation of recent results. There is a relatively large spread in the derived values of the SFR volume density - greater than a factor of two, beyond the errors quoted on the individual measurements. This is discussed briefly by \cite{2002MNRAS.330..621S} (who derive their own 1.4 GHz-based value of $(21\pm5) \times 10^{-3}\;\mathrm{M}_{\sun}\; \mathrm{yr}^{-1}\; \mathrm{Mpc}^{-3}$), who attribute the discrepancy to a systematic underestimation of the extinction using the Balmer decrement in some emission line-based studies.

The total value of $\rho_\mathrm{SFR}$ can also be decomposed into `UV' and `IR' components, by integrating the value of $\psi  \,\Phi(\psi) $ derived from each component individually. Doing so leads to values of $\rho_\mathrm{SFR}(\mathrm{IR}) = 0.011\; \mathrm{M}_{\sun}\; \mathrm{yr}^{-1}\; \mathrm{Mpc}^{-3}$, and $\rho_\mathrm{SFR}(\mathrm{UV}) = 0.012\; \mathrm{M}_{\sun}\; \mathrm{yr}^{-1}\; \mathrm{Mpc}^{-3}$. The LVL contribution is $0.0007\; \mathrm{M}_{\sun}\; \mathrm{yr}^{-1}\; \mathrm{Mpc}^{-3}$. This is 47\%, 50\%, 3\% of the total for the IR, UV, and LVL components respectively. This result - that about half of the energy from the total cosmic star formation budget is re-processed by dust - is well known, and is in line with previous studies. \cite{Takeuchi:2005aa} found that of their derived total SFR volume density ($19 \times 10^{-3} \; \mathrm{M}_{\sun}\; \mathrm{yr}^{-1}\; \mathrm{Mpc}^{-3}$), 56\% was from dust-reprocessed emission.\footnote{See also \cite{2005ARA&A..43..727L}, and references therein.} For consistency (and because our statistical AGN removal involves some uncertainty), we have checked the value of $\rho_\mathrm{SFR}$ calculated from the sample \textit{without} the statistical correction for AGN contamination (as per \S3.1). As the correction is only significant at the upper end (beyond $\psi^*$), the value only changes slightly: without any AGN correction applied, we calculate $\rho_\mathrm{SFR} = (26\pm2.2) \times 10^{-3}\;\mathrm{M}_{\sun}\; \mathrm{yr}^{-1}\; \mathrm{Mpc}^{-3}$.

We may also compute the fraction of the local cosmic star formation rate density occurring in starburst environments. For the purposes of such an analysis, we define a starburst as a system forming stars at $\geq$10 M$_{\sun}\; \mathrm{yr}^{-1}$. Using the star formation rate density distribution, we can thus integrate from 10 M$_{\sun}\; \mathrm{yr}^{-1}$ to infinity:

\[
f_\mathrm{burst}= \frac{1}{\rho_\mathrm{SFR}}\; \int_{\psi_\mathrm{burst}}^\infty  d\psi \; \psi  \,\Phi(\psi)  
\]
\begin{figure}
 \centerline{\includegraphics[scale=0.55, clip=true, trim = 10mm 125mm 40mm 30mm]{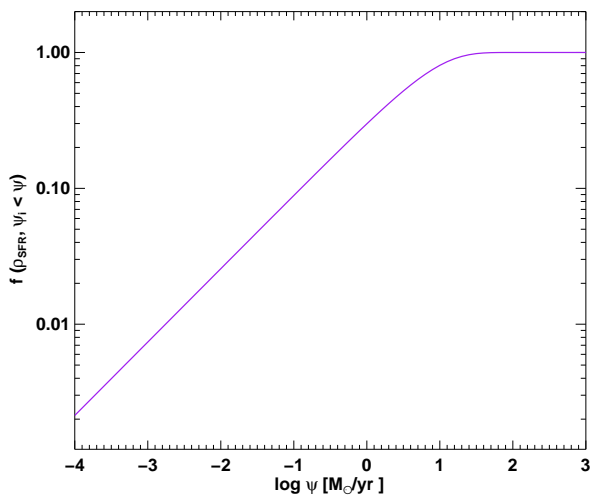}}
 \caption{The cumulative fraction of star formation rate volume density.}
 \label{fig:cum_sfr_den}
\end{figure}

For our data, this value is $0.0049\pm 0.00039$ M$_{\sun}\; \mathrm{yr}^{-1}\; \mathrm{Mpc}^{-3}$, or 20.4\% of the total star formation rate volume density; by our (admittedly somewhat crude) definition, one fifth of the star formation in the local Universe is provided by starbursts. This is consistent with the values found by \cite{2004MNRAS.351.1151B} using specific star formation rates from SDSS. Interestingly, \cite{2009ApJ...692.1305L} also find that 20\% of star formation in the dwarf galaxy population is concentrated in high H$\alpha$ equivalent width systems. 

There are many interesting values that can be derived from the distribution shown in Fig. \ref{fig:sfr_den}, including the starburst fraction (discussed above), the `dividing' SFR at which 50\% of the star formation is happening both above and below, and so on. Rather than providng a list of values for various integration limits, it is more enlightening to consider the behaviour of the cumulative fraction of star formation rate volume density, which is shown in Fig. \ref{fig:cum_sfr_den}. This shows SFR, plotted against the fraction of the total star formation volume density coming from SFRs \textit{lower} than that SFR. The data show a power-law increase in star formation rate volume density fraction, over 5 orders of magnitude until the truncation at $\sim20 \;\mathrm{M}_{\sun}\; \mathrm{yr}^{-1}$. From this it can be seen that the `50\%' divide occurs at $\sim 3\; \mathrm{M}_{\sun}\; \mathrm{yr}^{-1}$, about the SFR of the Milky Way (e.g. \citealt{2006A&A...459..113M}).

\begin{figure}
 \centerline{\includegraphics[scale=0.6, clip=true, trim = 10mm 125mm 40mm 30mm]{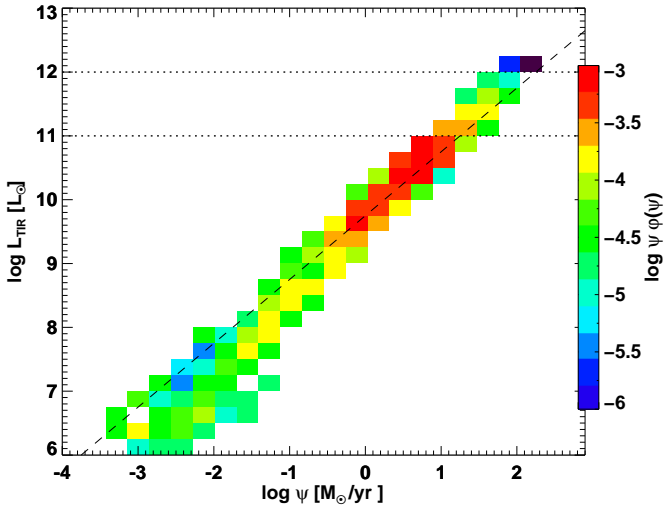}}
 \caption{The star formation rate volume density function $\psi\,\Phi(\psi)$ (as in Fig. 5) further expanded in a second dimension (along the ordinate) to show the breakdown with TIR luminosity. The colour-coded `z' axis represents $\psi\,\Phi(\psi)$. Horizontal lines are drawn at log(L)=\{11, 12\}, the respective defined minimum luminosities for LIRGs and ULIRGs. The dotted line shows the L(FIR)-SFR scaling given by Iglesias-P\'{a}ramo et al. (2006)}
 \label{fig:ULIRG}
\end{figure}

It is also interesting to consider the contribution to the total star formation rate volume density from LIRGs and ULIRGs. These IR-bright galaxies (defined as having $\mathrm{L}_{\mathrm{IR}} > 10^{11}\;\mathrm{L}_{\sun}$ and $> 10^{12}\;\mathrm{L}_{\sun}$ respectively) are rare in the local Universe, but become more and more important with lookback time, becoming an increasingly dominant component of the total star formation rate volume density at higher redshifts (\citealt{2005ApJ...619L..47S}; \citealt{2009A&A...496...57M}; \citealt{2010arXiv1008.0859G}). 

Fig. \ref{fig:ULIRG} shows the star formation rate distribution function shown in Fig. \ref{fig:sfr_den}, expanded along a second dimension with TIR luminosity (the colour-coded `z' axis corresponds to the value of $\psi\,\Phi(\psi)$). Horizontal lines have been drawn at the two characteristic luminosity cuts for LIRGs and ULIRGs, to illustrate the total contribution to the local SFRD coming from those galaxies. By integrating the SFRD function above and below the cutoff lines, we can estimate the contribution to the total from both LIRGs and ULIRGs; $9 \pm 1\%$ of total star formation is occurring in LIRGs, while just $0.6 \pm 0.2\%$ is occurring in ULIRGs. This is in good agreement with literature values - \cite{2010arXiv1008.0859G}, for example, reach similarly small estimates of $7\pm1$\% for LIRGs, and $0.4\pm0.1$\% for ULIRGs. As discussed above, the dominant object type driving the total star formation in the local Universe are normal, secularly evolving galaxies with star formation rates comparable to the Milky Way - despite their prodigious star formation rates, the sparsity of LIRGs/ULIRGs means that they do not contribute significantly. 

Due to the (U)LIRGs' bright IR luminosities, the derived LIRG/ULIRG fractions are highly sensitive to the nature of the AGN correction used, which strongly affects the behaviour of the LF at the bright end. If we do not correct for AGN contamination as per \S 3.1, the fractional contribution to the total SFRD from LIRGs and ULIRGs respectively is $14 \pm 2\%$ and $1.5 \pm 0.4\%$. It should be noted, then, that our original derived (U)LIRG contributions are highly dependent on the AGN correction.

\subsection{Dust obscured star formation}
\label{sec:irx}

\begin{figure}
 \centerline{\includegraphics[scale=0.55, clip=true, trim = 10mm 125mm 40mm 30mm]{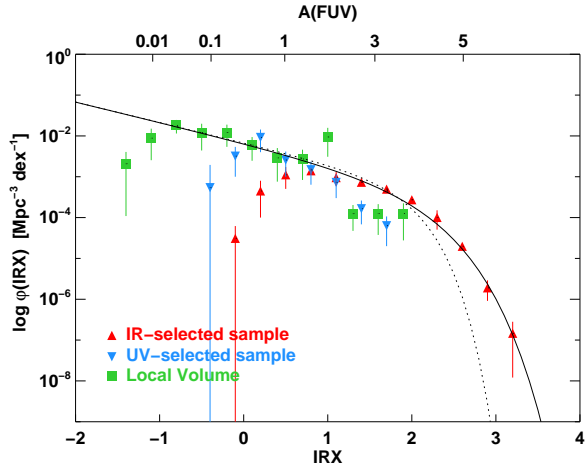}}
 \caption{The distribution function of IRX for the three samples. The red, blue, and green lines are, respectively, the IR-selected, UV-selected, and LVL datasets. The solid black line is the best-fitting modified Schechter fit (see text), and the dotted black line shows the same parameters converted into a standard Schechter fit by setting the high-end slope $\beta=1$. It can be seen that a normal Schechter function is a poor fit to the data. The equivalent FUV attenuation has been shown on the upper x-axis, calculated as detailed in the text. }
 \label{fig:IRX_dist}
\end{figure}

It is possible to examine the distribution function of extinction, in much the same way as we have previously examined the distribution function of luminosity and star formation. This will examine the global behaviour of dust obscured star formation, as a function of the amount of obscuration. We use as our measure of `extinction' or `dust obscuration' the ratio of IR to observed UV luminosities, IRX, as described above in \S\ref{sec:sfr}. Having a strong positive correlation with both luminosity and star formation rate (\citealt{1996ApJ...457..645W}; \citealt{2001AJ....122..288H}; \citealt{2005ApJ...619L..51B}; \citealt{2010A&A...514A...4T}) the IRX distribution function should resemble the Schechter-like distribution functions derived elsewhere in this work. 

Fig. \ref{fig:IRX_dist} shows the distribution functions of IRX for the three samples included in this work. The three samples have been shown separately, as a `resultant' sample, constructed by combining the different samples with different selection functions has little physical interpretation in this case; the distribution of IRX, and the difference between the samples, sheds light on FIR and FUV selection effects \citep{2006ApJ...646..834X} as much as it does the underlying physical properties of the galaxies in question. 

It is important to note here that, in contrast to the work with luminosity and SFR described above, the choice to include/exclude non-detections significantly alters the shape of the IRX distribution function, simply because the IRX is defined as a ratio of luminosities. So, while a non-detection in the UV has little effect in terms of the SFR function (it simply implies that the bolometric output is predominantly in the IR, and the UV is negligible), it has the result of making the value of IRX formally infinite. If non-detections are assumed to have flux equal to the limiting flux (as above), the IRX distribution function becomes essentially meaningless, as it is driven entirely by the extreme upper/lower limit values of IRX assigned semi-arbitrarily to the non-detections. We therefore chose to \textit{exclude} non-detections for the purpose of this IRX analysis. The number of galaxies thus excluded from the IR and UV samples is, respectively, 1784 and 73.  As a result of excluding the non-detections, the distribution functions shown in Fig. \ref{fig:IRX_dist} can be interpreted as lower bounds on the IRX distribution functions.

All three samples exhibit broadly similar behaviour in their value of $\Phi$(IRX): a decrease in $\Phi$ towards higher values of IRX (analogous to the decrease in $\Phi$(L) for rare high-luminosity systems), and a second dip at the lowest values of IRX probed. With the exception of an overdensity at IRX=1 in the LVL sample, the three samples do share a common Schechter-like `envelope', though the individual samples do differ. The UV and LVL samples fall off steeply between an IRX of 1 and 2. In contrast, the high-IRX end (comprised entirely of the IR-selected sample) is much shallower than would be expected from a standard Schechter function, however, and is best fit with a modified Schechter function, which has been adapted following \cite{2010MNRAS.402.1693H} to leave the high-IRX end slope $\beta$ as a free parameter:

\[
\Phi(X) dX = \Phi^* \left(\frac{X}{X^*}\right)^{\alpha} exp[-(X/X^*)^{\beta}] d\left(\frac{X}{X^*}\right),
\]
where a high end slope $\beta < 1$ allows for a shallower fall-off at high IRX than a standard Schechter function ($\beta = 1$). 

This is plotted on Fig. \ref{fig:IRX_dist} as a solid black line (the standard Schechter fit using the same parameters is, for comparison, shown as a dotted line). This has a low-end slope of $\alpha = -1.51 \pm 0.11$, and a high-end slope $\beta = 0.63 \pm 0.09$. This is significantly shallower than 1, which would be expected given the previously-noted strong correlation between luminosity/SFR and IRX, and deserves further examination. 

It is clear from Fig. \ref{fig:IRX_dist} that the IR-selected sample contains more high-IRX objects than the UV-selected sample. This is to be expected: galaxies having a high IRX (i.e. being IR-bright, UV-faint) are more
likely to be over-represented in an IR-selected sample compared to a
UV-selected sample, if only because the effective volume probed in the
IR is far larger than in the UV for such an object. (The converse is
true too, of course; low-IRX things which are UV bright and IR faint
have larger effective volumes probed by a UV survey, so we would
expect a UV survey to contain disproportionately many low-IRX
galaxies).

\begin{figure}
 \centerline{\includegraphics[scale=0.065]{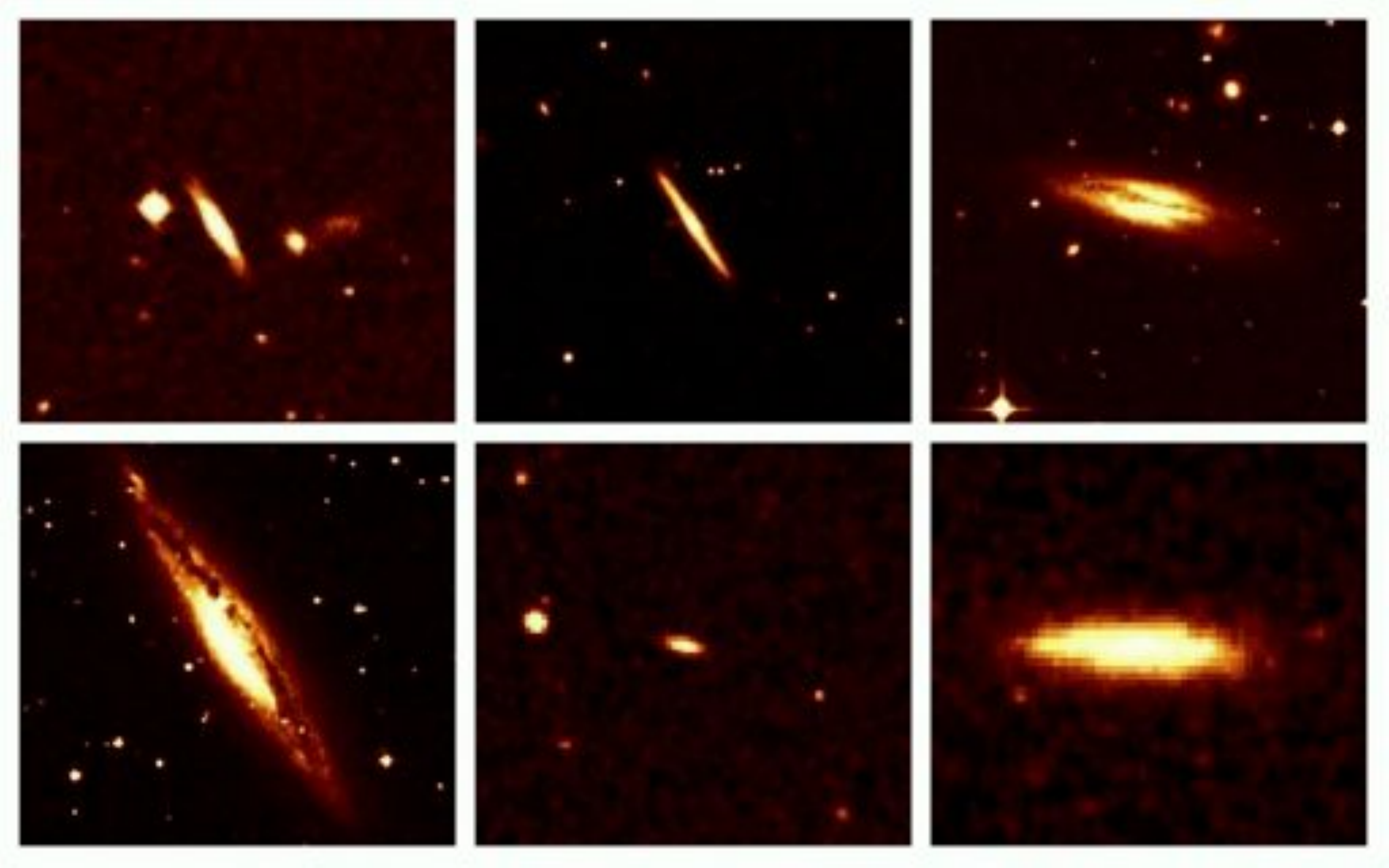}}
 \caption{A selection of the high-IRX galaxies from the IR-selected sample, showing their unusual/edge-on morphology and prominent dust lanes. Clockwise from the top left: IRAS F14299+3631; F01221+0944; F21596-1909; F14263+2555; F13228+1837; F11290-3001. Images obtained from NASA NED.}
 \label{fig:galaxy_pics}
\end{figure}

Using NED, we looked up the high-IRX galaxies in our sample, which make up the top of the distribution. The galaxies inhabiting this extended shallow `tail' to the distribution are, predominantly, high-inclination discs (see Fig. \ref{fig:galaxy_pics} for a selection). We suggest that this bias towards inclined galaxies (which would have boosted IRXs due to the greater amount of material obscuring the star forming disc/bulge) is a possible explanation for the shallowness of the distribution. If this is the case, then the shallow high-end slope seems to be driven by inclination (and its associated radiative transfer effects) as much as inherent, angle-averaged extinction. 


This deviation from a Schechter function is, to some extent, expected. Press-Schechter formalism describes the distribution of DM halo masses -- parameters that are well correlated with halo mass should have a Schechter-like scale distribution. As the parameter in question becomes further removed from the underlying halo mass, there is less reason to expect that a Schechter function will be a good model of the parameter distribution. The additional complication of complex, inclination-dependent radiative transfer effects would cause the IRX distribution to take a non-Schechter form.

\begin{figure}
 \centerline{\includegraphics[scale=0.6, clip=true, trim = 10mm 125mm 40mm 30mm]{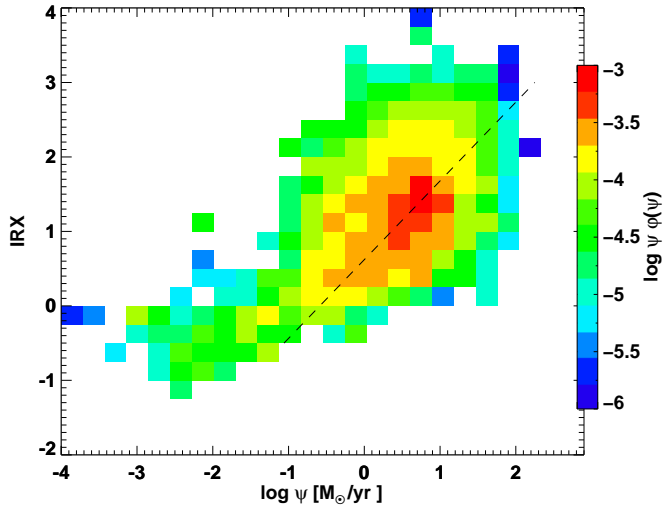}}
 \caption{The star formation rate volume density function as a function of SFR (Fig. 5), further expanded in a second dimension (along the ordinate) to show the breakdown with IRX. The `z' axis represents $\psi \,\Phi(\psi)$. The dotted line shows the IRX-SFR relationship derived from the L$_{\mathrm{IR}}$ vs. E(B -- V) relationship given by Hopkins et al. (2001). E(B -- V) was converted into $A_{\mathrm{FUV}}$ using the Cardelli (1989) extinction law with $R_V$ = 3.1, which gives $A_{\mathrm{FUV}}$ = 8.0 E(B -- V). IRX and $\psi$ were obtained from $A_{\mathrm{FUV}}$ and L$_{\mathrm{IR}}$ as above.}
 \label{fig:IRX_sfrd}
\end{figure}

It is again possible to break down the star formation rate volume density distribution (shown in Fig. \ref{fig:sfr_den}) with IRX, to see at which extinction level the contribution is greatest. This is shown in Fig. \ref{fig:IRX_sfrd}. The SFR volume density function is decomposed into bins of IRX along the y axis, while the colour-coded `z' axis corresponds to $\psi\,\Phi(\psi)$.  The peak in the SFRD function which occurs at $\psi \sim 3 \;\mathrm{M}_{\sun}\; \mathrm{yr}^{-1}$, occurs at IRX $\sim1.4$ in this breakdown. While the SFRD function decreases sharply towards higher SFRs, however, the decline towards higher values of IRX is much more shallow, with a non-trivial fraction of the total SFRD coming from highly extincted systems. $13\pm1$\% of the total star formation rate volume density comes from systems with IRX$>2$, dusty systems with a factor of 100 difference between their IR and UV luminosities.

\nocite{Hopkins:2001aa}
\nocite{Cardelli:1989aa}

\section{Conclusions}
We have presented an analysis of the IR- and UV-selected luminosity functions for galaxies in the $z\sim0$ Universe. We make use of large survey datasets ($>10,000$ galaxies) from the IRAS and GALEX satellites to provide us with good number statistics; we also utilise the Local Volume Legacy survey to extend the field relations 1-2 orders of magnitude deeper than before, probing the faintest visible star forming galaxies. Our main conclusions are as follows:

\begin{itemize}
\item{The distribution function of star formation in the local Universe has a faint end slope of $\alpha = -1.51\pm0.07$. This appears to be constant and monotonic as far back as the data allow us to go, down to the faintest dwarf galaxies with star formation rates $< 0.001 \;\mathrm{M}_{\sun}\; \mathrm{yr}^{-1}$.}\\

\item{The star formation rate volume density of the local ($z\sim0$) Universe is found to be $0.025 \pm 0.0016\;\mathrm{M}_{\sun}\; \mathrm{yr}^{-1} \mathrm{Mpc}^{-3}$. The distribution function of star formation rate volume density shows that this value is predominantly driven by modest star forming galaxies, with star formation rates of $\sim 3 \;\mathrm{M}_{\sun}\; \mathrm{yr}^{-1}$, similar to the Milky Way. `Starburst' galaxies, with SFRs $\geq10 \;\mathrm{M}_{\sun}\; \mathrm{yr}^{-1}$, contribute 20\% to the local cosmic star formation rate density.}\\

\item{In contrast to the high-$z$ Universe, LIRGs and ULIRGs play only a modest role in determining the total star formation rate volume density, contributing $9\pm1 \%$ and $0.7\pm 0.2\%$ respectively.}\\

\item{The distribution function of  ratio $\mathrm{L}_{\mathrm{IR}} / \mathrm{L}_{\mathrm{FUV}}$, `IRX', behaves somewhat differently to that of star formation, despite the strong correlation. The low-IRX slope is comparable, but the high-IRX slope is significantly shallower, and is very poorly fit by a standard Schechter function. This is partly attributable to the addition of geometric effects (whereby edge-on galaxies have inflated extinctions relative to their face-on counterparts), and vary the bright end slope $\beta$ to the data provides a good fit for $\beta = 0.63 \pm 0.09$.}\\

\item{Breaking down $\psi\,\Phi(\psi)$ with IRX, it can be seen that the peak of SFRD occurs at IRX $\sim 1.4$. Extremely dusty systems contribute a minor but significant fraction to the total SFRD, with $13\pm1$\% coming from dusty galaxies with IRX$>2$.  }

\end{itemize}

\section*{Acknowledgments}
We would like to thank Matt Malkan for helpful and enlightening discussions. This research has made use of the NASA/IPAC Extragalactic Database (NED) which is operated by the Jet Propulsion Laboratory, California Institute of Technology, under contract with the National Aeronautics and Space Administration. We acknowledge the usage of the HyperLeda database (http://leda.univ-lyon1.fr). MB is supported by STFC.

\bibliography{/Users/Matt/Documents/mybib}{}
\bibliographystyle{mn2e}

\label{lastpage}

\end{document}